\documentclass[12pt]{article}
\usepackage{amsmath,amssymb,amsfonts,amsthm}
\title{Quantum statistical properties of some new classes of intelligent states associated with special quantum      systems}
\author{M. K. Tavassoly  and A. Parsaiean
\\
\footnotesize{Atomic and Molecular Group, Faculty  of Physics, Yazd University, Yazd, Iran}
\\ \footnotesize{e-mail: mktavassoly@yazduni.ac.ir  } }

\begin{document}

\newcommand{\norm}[1]{\left\Vert#1\right\Vert}
\newcommand{\abs}[1]{\left\vert#1\right\vert}
\newcommand{\set}[1]{\left\{#1\right\}}
\newcommand{\R}{\mathbb R}
\newcommand{\I}{\mathbb{I}}
\newcommand{\C}{\mathbb C}
\newcommand{\eps}{\varepsilon}
\newcommand{\To}{\longrightarrow}
\newcommand{\BX}{\mathbf{B}(X)}
\newcommand{\HH}{\mathfrak{H}}
\newcommand{\D}{\mathcal{D}}
\newcommand{\N}{\mathcal{N}}
\newcommand{\la}{\lambda}
\newcommand{\af}{a^{ }_F}
\newcommand{\afd}{a^\dag_F}
\newcommand{\afy}{a^{ }_{F^{-1}}}
\newcommand{\afdy}{a^\dag_{F^{-1}}}
\newcommand{\fn}{\phi^{ }_n}
 \newcommand{\HD}{\hat{\mathcal{H}}}

 \maketitle

\begin{abstract}
Based on the {\it nonlinear coherent states} method, a general and simple algebraic
formalism for the construction of \textit{`$f$-deformed intelligent states'} has been
introduced. The structure has the potentiality to apply to systems with a
known discrete spectrum as well as the generalized coherent states with known
nonlinearity function $f (n)$. As some physical appearance of the proposed
formalism, a few new classes of intelligent states associated with \textit{`center of-
mass motion of a trapped ion'}, \textit{`harmonious states'} and \textit{`hydrogen-like
spectrum'} have been realized. Finally, the nonclassicality of the obtained
states has been investigated. To achieve this purpose the quantum statistical
properties using the Mandel parameter and the squeezing of the quadratures of
the radiation field corresponding to the introduced states have been established
numerically.
\end{abstract}



\section{Introduction}\label{sec-intro}
Nowadays \textit{`coherent states'} first introduced by Schr\"{o}dinger [29] are known as venerable objects in quantum physics, especially in quantum optics [1]. Generalizations of these states led to new terms as nonlinear coherent states, squeezed states and intelligent states corresponding to different quantum systems [4-6, 9, 31]. Among them, the terminology of intelligent states was first introduced by the original works of Aragone et al [4, 5] and then these states have been developed by others (for example see [6, 31]). The most interest and motivation in these states, as all states of the quantized electromagnetic radiation field, have been discussed in quantum optics are \textit{nonclassical} properties such as sub-Poissonian statistics and/or squeezing in one of the quadratures of the field. Nonclassical states may be used in various ways in tuning measurement resolution beyond the standard limit. We may refer to the applications of the squeezed states of light in high-precision modern experiments of physics, accurate  measurements, optics communications and detection of weak signals due to the reduction of quantum noise as prominent examples [15, 32]. For a squeezed review on the nonclassical states and the usefulness and applications of these states see [8] and references
therein. On the other hand, the \textit{nonlinear coherent state} is an interesting notion in the quantum optics field in recent years characterized by an operator-valued function $f (n)$ [16, 17]. Our first aim in the present paper is to propose a simple and general algebraic method providing the description of the intelligent states' formalism in terms of the nonlinearity function associated with any nonlinear coherent states. As we will observe, this link is also possible between any quantum system with known discrete spectrum and intelligent states. Secondly, the nonclassicality of a few new intelligent states constructed by the presented formalism will be established. We briefly recall the basic structure of the intelligent states. Suppose $A$ and $B$ are two self-adjoint operators with commutator $[A, B]$. Intelligent states are understood to be the states satisfying the equality sign in the Heisenberg uncertainty relation
\begin{equation}\label{agh}
\Delta A \Delta B \geqslant \frac 1 2 | \langle [A, B] \rangle |
\end{equation}
where
\begin{equation}\label{va}
\Delta A = \sqrt{\langle A ^2 \rangle - \langle A \rangle ^2},
\end{equation}
and similarly for $\Delta B$. We will concerned with states $ | \Psi \rangle $ which are not the eigenstates of either A or B, and which are intelligent. By some manipulations it is proved that these states are the solutions of the eigenvalue equation [18]
\begin{equation}\label{kt}
(A + i\lambda B)| \Psi \rangle = z | \Psi \rangle,
\end{equation}
where $\lambda$ is complex for the generalized intelligent states and real for the ordinary ones and the eigenvalue $z \in \C$. In the present paper, we restrict ourselves to the case $\lambda \in \R $. The physical role of $\lambda$ parameter in (3) in controlling the squeezing of $ | \Psi \rangle $ will be clarified in the
next section.

   As the simplest example of these states one may set $A \equiv x$ and $B \equiv p$ , i.e. the dimensionless position and momentum operators defined as
\begin{equation}\label{xp}
   x = \frac{a + a ^ \dag}{\sqrt{2}},   \qquad  p =  \frac{a - a ^ \dag}{\sqrt{2} i},
\end{equation}
   in terms of $a$, $a^ \dagger$ as the bosonic annihilation, creation operators, respectively, with $[x,p]=i$. Continuing the calculation with $\lambda =1$ in the eigenvalue equation (3) the elementary intelligent state is obtained as the \textit{standard coherent states} of the quantum harmonic oscillator, for which we have $ \Delta x \Delta p = \frac{1}{2}$:

 \begin{equation}\label{sy}
    |\Psi \rangle \equiv |\alpha \rangle = e^{-| \alpha |^2/2}
    \sum_{n=0}^{\infty}\frac{\alpha^{n}}{\sqrt{
    n!}}|n\rangle,
\end{equation}
   where $\alpha \equiv z/\sqrt{2} \in \C$. In addition, another well-known example is the intelligent states of $SU(1, 1)$ group [6, 11]. For constructing these states one may   set $A\equiv K_1=\frac{1}{2}(K_+ + K_-)$ and $ B\equiv K_1=\frac{1}{2}(K_+ - K_-)$ in (3) where $\{K_+, K_-, K_0\}$ are the generators of the $su(1,1)$ Lie algebra with commutators $[K_+, K_-]=-2K_0$ and $[K_0, K_\pm]=\pm K_\pm$. The role of the intelligent states of $SU(1, 1)$ (and $SU(2)$ group) in increasing the sensitivity of the
interferometric measurement has been studied in [22]. Also the generalized intelligent states of $SU(N)$ algebra have been introduced in [7]. Anyway, the generalization of the intelligent states may be considered along the generalization of the coherent states with the same motivations in quantum optics.

   The content of the paper is as follows: first, we will briefly review the nonlinear coherent state method and demonstrate the place of this formalism in the construction of the generalized intelligent states in section 2. Next, in section 3 by defining the two Hermitian operators X as the deformed position and P as the deformed momentum operator in terms of the nonlinearity function $f(n)$, the general form of the (\textit{$f$-deformed}) intelligent states in the language of nonlinearity function has been introduced. Finally, in section 4 the presented formalism has been applied to a few well-known quantum systems \textit{"center-of-mass motion of a trapped ion"},
 \textit{"harmonious states"} and \textit{"hydrogen-like spectrum"}, and at the same time the nonclassicality
 of the corresponding intelligent states has been justified numerically. To the best of
 our knowledge the intelligent states associated with these latter systems have not
 been appeared in the literature.

\section{Intelligent states associated with nonlinear coherent states}\label{lklk}

   In this section, we want to formulate the general formalism of the intelligent states briefly
reviewed in section 1 in a new algebraic structure. The \textit{"nonlinear coherent state"}, which is a notion attracted much attention in recent decade, provides a powerful technique to achieve this purpose. The usefulness of this method to obtain the ladder operators corresponding to any solvable quantum system with discrete spectrum and also in the classification of the generalized coherent states has been frequently discussed [24, 25]. Recall that the single-mode nonlinear coherent states are essentially based on the deformation of bosonic annihilation and creation operators according to the relations [16, 17]

\begin{equation}\label{af}
 A= af(\hat{n})=f(\hat{n}+1)a,
\end{equation}
\begin{equation}\label{adf}
A^\dagger =f(\hat{n})a^\dagger =a^\dagger f(\hat{n}+1),
\end{equation}
with actions on the Fock space as
\begin{equation}\label{ak2tn}
   A | n\rangle =\sqrt{n} f(n) | n-1 \rangle,
\end{equation}
\begin{equation}\label{aktn}
   A^\dagger | n\rangle =\sqrt{n+1} f(n+1) | n+1 \rangle,
\end{equation}
where $f (n)$ is an operator-valued function of number operator $n = a^\dagger a$ responsible for the
nonlinearity of the generalized coherent states. Generally, $f (n)$ is a real- or complex-valued
function but in this paper we deal only with the real-valued ones (we may recall that for
Gazeau$-$Klauder coherent states $f_{GK}(n)$ is a complex-valued function [25]). The single-mode
nonlinear coherent states defined as the eigenstate of the deformed annihilation operator are
obtained as follows:
\begin{equation}\label{zf1}
| z,f \rangle = N(| z |^2)^{-1/2} \sum_{n=0}^{\infty}C_n z^n|n\rangle=N(| z |^2)^{-1/2} \sum_{n=0}^{\infty}\frac{z^n}{\sqrt{[nf^2(n)]!}}|n\rangle,
\end{equation}
where $z \in \C$. By definition $C_0 = 1$,$ [f (n)]! \doteq
f (n)f (n -1) \cdots f (1) $and the normalization
constant is determined as
 \begin{equation}\label{normal}
 N(| z |^2)= \sum_{n=0}^{\infty}\frac{| z |^{2n}}{[nf^2(n)]!}.
 \end{equation}
 Inversely, knowing coefficients $ C_n $ the function $f (n)$ may be found by
 \begin{equation}\label{fn1}
    f(n)=\frac{C_{n-1}}{\sqrt{n}C_n}.
\end{equation}
    A lot of analysis of the generalized coherent states to establish their nonlinearity nature and to introduce the corresponding nonlinearity function and ladder operators has been previously done by Roknizadeh et al in [24, 25]. To say a few, one may refer to the $SU(1, 1)$ coherent states [21, 23], center-of-mass motion of trapped ion [17], hydrogen-like spectrum [13], infinite square well and P\"{o}schl$-$Teller potential [3], PS (Penson$-$Solomon [20]) and KPS (Klauder$-$Penson$-$Sixdeniers [14]) nonlinear coherent states.

      Now one can define the \textit{"generalized ($f$-deformed) position and momentum operators"}
       using the $f$-deformed ladder operators in (6) and (7) as follows:

  \begin{equation}\label{xptarif}
     X = \frac{A + A ^ \dag}{\sqrt{2}},   \qquad  P =  \frac{A - A ^ \dag}{\sqrt{2} i},
  \end{equation}
  which are clearly two Hermitian operators. The commutator [X, P] can be found as
   \begin{equation}\label{jabjaxp}
     [X, P]=i[(n+1)f^2(n+1)-nf^2(n)].
  \end{equation}
Subsequently, the intelligent states may be defined as the states which satisfy the equal sign in the Heisenberg uncertainty relation
\begin{equation}\label{ghateat}
  \Delta X \Delta P \geqslant \frac{1}{2} | \langle [ X, P ]\rangle |.
\end{equation}

Therefore, the equivalent requirement for a state to be intelligent is to satisfy the \textit{"basic eigenvalue problem"}:
\begin{equation}\label{ilz}
(X+i\lambda P) | z,\lambda \rangle =\sqrt{2} z | z,\lambda \rangle,
\end{equation}
where $\lambda \in \R $ and $z\in \C$. The above explanations clearly led one to a simple algebraic method which allows one to construct new classes of the intelligent states corresponding to any nonlinear coherent states. This work that is a part of the purpose of the present paper will be possible, only if the associated nonlinearity function $f (n)$ and so rising and lowering operators were discovered.

It is proved that the $\lambda$ parameter in (3) and (16) is determined by $| \lambda |=\frac{\Delta A}{\Delta B}$ and $| \lambda |=\frac{\Delta X}{\Delta P}$ ,
respectively [6]. Therefore, the role of this parameter to control the squeezing effect inherent in the states $| z, \lambda \rangle$ will be obvious. Since for the intelligent states introduced in this paper one concerns with the equality sign in (1) it is seen that for $\lambda=1$ and $\lambda \neq 1$ the obtained states will
be (generalized) coherent states and squeezed states, respectively. Precisely speaking, when $| \lambda |<1$ the intelligent states are squeezed in $A$ (or $X$) and when $| \lambda |>1$ the intelligent states are squeezed in $B$ (or $P$). When $| \lambda |=1$, i.e. for which equal fluctuations in the deformed position and momentum operators $X, P$ occur, the states are named (generalized) coherent states. These sets of states are also the already known (generalized) coherent states obtained through the \textit{"dynamical"} generalization of the coherent states, i.e. the coherent states for potentials
other than harmonic oscillator [19]. In this regard, these sets of the intelligent states are along with the searching line of Shr\"{o}dinger who was assessing the most classical-like quantum states when he first introduced the coherent states. The latter set of the intelligent states ($\lambda =1$)
possesses similar property, but in the deformed space $X, P$.

\section{Solutions of the basic eigenvalue equation of intelligent states}\label{sddddd}

   Now, we return again to the basic eigenvalue equation in (16) which can be converted in the following form using (6) and (7):
   \begin{equation}\label{sop1}
[(1-\lambda )A^\dagger +(1+\lambda)A]| z, \lambda \rangle =2z  | z, \lambda \rangle.
\end{equation}
In this stage, setting
\begin{equation}\label{sop2}
| z, \lambda \rangle = \sum_{n=0}^{\infty}c_n | n \rangle
\end{equation}
in (17) it can be solved for different values of $\lambda$ and z. In all four cases which follow the task is to find the appropriate $c_n$ coefficients after summation sign on the rhs of equation (18):\\
\begin{description}
    \item[(i)] $\lambda=1$; $z\neq0$. In this case, equation (17) reduces to the simple relation $A| z\rangle=z | z\rangle$, the states satisfying the latter equation are $\lambda$-independent and they are indeed the well-known nonlinear coherent states $| z, f\rangle$ in (10) discussed frequently in the literature [16, 17]. In
this sense, along with the standard coherent states, the nonlinear coherent states introduced in (10) are also intelligent. In addition, we can include the dual of the nonlinear coherent states with $ \widetilde{f}(n)=\frac{1}{f(n)}$ [2, 25, 26] in this category.

   \item[(ii)] $\lambda=-1$; $z\neq0$. With these parameters we get the eigenvalue equation $A^\dagger| z\rangle=z | z\rangle$ with no solution. Precisely speaking, all of the coefficients $c_n$ are zero showing the already
known result that there is no quantum state which is the eigenvalue of the generalized
creation operator.

 \item[(iii)] $\lambda \neq -1$; $z=0$. In this case, solving the eigenvalue problem (17) gives all of odd $c_n$'s
zero, while the even $c_n$'s are obtained as follows:

\begin{equation}\label{So1}
c_{2n}=c_0\left(\frac{\lambda -1}{\lambda+1}\right)^n \sqrt{\frac{(2n-1)!!}{(2n)!!}} \frac{[f(2n-1)]!!}{[f(2n)]!!},
\end{equation}
where the coefficients $c_0$ may be determined using the normalization constraint as

\begin{equation}\label{fact}
c_0 =\left[\sum_{n=0}^{\infty}\left(\frac{\lambda -1}{\lambda+1}\right)^{2n} \frac{(2n-1)!!}{(2n)!!} \left(\frac{[f(2n-1)]!!}{[f(2n)]!!}\right)^2\right]^{-1/2}.
\end{equation}

 \item[(iv)] $\lambda \neq -1$; $z\neq0$. Finally, in this case for the coefficients $c_n$'s one gets

\begin{equation}\label{SS11}
c_n=\frac{c_0}{\sqrt{n!}[f(n)]!}\left(\frac{2z}{1+\lambda}\right)^n \left[1+\sum_{h=1}^{[n/2]}(-1)^h\frac{(1-\lambda ^2)^h}{(2z)^{2h}}S(n, h)\right],
\end{equation}
  where in (21) $[n/2]$ represents the integer part of $n/2$ and the function $S(n, h)$ is defined by

\begin{equation}\label{norm03}
S(n,h)=\sum_{m_1=1}^{n-(2h-1)}m_1 f^2(m_1)\left[\sum_{m_2=m_1+2}^{n-(2h-3)}m_2 f^2(m_2)\left[\cdots \left[\sum_{m_h=m_{h-1}+2}^{n-1}m_h f^2(m_h)\right]\cdots\right]\right],
\end{equation}
and $c_0$ may be determined from the normalization condition as follows:

\begin{equation}\label{DSS-Kwek}
c_0=\left[\sum_{n=0}^{\infty}\left| \frac{2z}{\lambda+1} \right|^{2n} (\sqrt{n!}[f(n)]!)^{-2}\left(1+\sum_{h=1}^{[n/2]}(-1)^h\frac{(1-\lambda^2)^h}{(2z)^{(2h)}}S(n,h)\right)^2\right]^{-1/2}.
\end{equation}
\end{description}

  It must be notified that a formalism for the construction of the generalized intelligent states for exactly solvable quantum systems has also been proposed previously in [9] via super-symmetric approach. While our presentation is based on an algebraic technique in
terms of nonlinearity function of the nonlinear coherent states, which seems to be simpler, at
the same time it suggests a powerful technique to use this structure to any quantum systems with  known nonlinear coherent states. The main step to start the procedure is to know the  nonlinearity function responsible for the system. It can be recognized that the presented
structure would be useful whether the nonlinearity function itself is determined, otherwise the corresponding discrete spectrum $e_n$ is known and so through the relation $e_n = nf^2(n)$ the nonlinearity function may be explored [24, 25].\\

\section{ Nonclassical properties of the introduced intelligent states associated with special quantum systems}\label{ggg}

In this section, we want to apply the proposed formalism to the
quantum systems describing the center-of-mass motion of trapped
ion, harmonious states and the hydrogen-like spectrum, all of
which are the corresponding coherent states, and the nonlinearity
natures have been already discussed in the literature. So, after
introducing the associated intelligent states, have not appeared
in the literature up to now, we shall establish some of the their
nonclassical features.

The nonclassical states do not have any classical analogue.
Generally, a state is known as a nonclassical state if the
Glauber$-$Sudarshan $P(\alpha)$ function [12, 30] cannot be
interpreted as a probability density. However, in practice one
cannot directly apply this criterion, since in general this
function is strongly singular and thus this definition may hardly
be applied for investigating the nonclassicality of a state [28].
Altogether, this purpose has been achieved commonly by checking
the \textit{squeezing effect, sub-Poissonian statistics and
oscillatory number distribution}.

A common feature of all the above-mentioned criteria is that the
corresponding P-function of a nonclassical states is not positive
definite. It is worth paying attention to the fact that each of
the above effects (squeezing and sub-Poissonian which we will use
in the this paper) is a sufficient condition for a state to
exhibit nonclassicality. The quantum states of light in the
cavity are customarily characterized by the Mandel parameter $Q$,
which is related to the fluctuations of intensity operator $n$
through

 \begin{equation}\label{S0S22}
 Q = \frac{\langle (\Delta n)^2\rangle - \langle n\rangle}{\langle n\rangle},
 \end{equation}
  where $(\Delta n)=\sqrt{\langle n^2 \rangle -\langle n\rangle^2}$ and $n = a^\dagger a$ is the number of photons. The $Q$
quantity vanishes for \textit{`standard coherent states'}
(Poissonian), positive for \textit{`classical'}
(super-Poissonian) and negative for \textit{`nonclassical'}
(sub-Poissonian) light. Antibunching also has been defined as a
phenomenon in which the fluctuations in photon number reduced
below the Poissonian level $((\Delta n)^2 < \langle n\rangle)$.
Sub-Poissonian and antibunching of photons occur simultaneously
at least in the fist order of these effects, which is the case we
have considered in the present paper. The next condition that a
nonclassical state may possess is the squeezing in one of the
quadratures of the quantized radiation field. Based on the
definitions in (4) the uncertainty in coordinate and momentum
operator will be obtained as follows:
\begin{eqnarray}\label{norm20p}
(\Delta x)^2=\langle x^2\rangle-\langle x\rangle ^2=0.5[1+\langle a^2\rangle+\langle a^{\dagger2}\rangle+2\langle a^\dagger a \rangle-\langle a\rangle^2-\langle a^\dagger \rangle^2-2\langle a\rangle \langle a^\dagger \rangle],
 \end{eqnarray}\begin{eqnarray}\label{norm20p}
(\Delta p)^2=\langle p^2\rangle-\langle p\rangle ^2=0.5[1-\langle a^2\rangle-\langle a^{\dagger2}\rangle+2\langle a^\dagger a \rangle+\langle a\rangle^2+\langle a^\dagger \rangle^2-2\langle a\rangle \langle a^\dagger \rangle].\nonumber
 \end{eqnarray}
For the investigation of the squeezing of the quadratures we have
used the $q_i$ parameter defined by
  \begin{equation}\label{llll}
   q_i=\frac{\langle (\Delta q_i)^2\rangle -0.5}{0.5},
  \end{equation}
where $q_1\equiv x$ and $q_2\equiv p$. The state is squeezed in
$q_i$ quadrature if the inequality $-1<q_i<0$ is satisfied.
Obviously, in the continuation of the paper we have considered
the (dimensionless) momentum and position ($x$, $p$) in (4) in
our calculations, since as we have pointed out at the end of
section 3 the occurrence of squeezing phenomenon in the
generalized momentum and position ($X, P$) in (4) for each set of
the intelligent states is specified by $\lambda$ parameter
entirely.

The obtained states in the present paper are more complicated to
be analyzed analytically, so the $Q$ and $q_i$ parameters may be
calculated by numerical computations. For this purpose,one has to
calculate the expectation values expressed in equations (24) and
(26) over the three classes of acceptable (normalizable)
intelligent states in consideration $| z, \lambda\rangle$
according to the items (i), (iii) and (iv) we have itemized in
section 3. Therefore, corresponding to any system with known
discrete spectrum (or a known function $f (n)$) there exist three
classes of the intelligent states, may be explicitly derived from
the three sets of equation (10) (item (i)), equation (19) (item
(iii)) and finally equation (21) (item (iv)). In the following
discussions, we will only refer to the items (i), (iii) and (iv)
to specify the special type of the intelligent states in
consideration. The oscillatory number distribution of the states
of (iii) type is an obvious matter, independent of the chosen
form of the function $f(n)$. Therefore, in the continuation of
the paper which we deal with special quantum systems such as
\textit{`center-of-mass motion of a trapped ion'},
\textit{`harmonious states'} and \textit{`hydrogen-like
spectrum'}, our calculations have been focused on the
investigation of squeezing phenomena and the behavior of the
Mandel parameter.\\

(1) \textit{Center-of-mass motion of a trapped ion}. Nonclassical
states of atomic center-of-mass motion of trapped ion have played
an important role in fundamentals of quantum mechanics and also in
experimental modern physics due to their important applications
in various fields of physics as well as in quantum engineering.
The nonlinearity function of this system is well known as

  \begin{equation}\label{DSS8555}
   f_{TI}(n)=\frac{L_{n}^{1}(\eta ^2)}{(n+1)L_{n}^{0}(\eta ^2)},
  \end{equation}
   where $\eta$ is Lamb$-$Dicke parameter and $L_{m}^{n}(x)$
 are the associated Laguerre polynomials. The trapped ion
nonlinear coherent states are defined on the whole complex plane.

Now we want to investigate the nonclassical properties of the
three classes of the intelligent states when $f_{TI}(n)$ is used
in formulations by numerical calculations. Figures 1(a) and (b)
concerned with the trapped ion states in item (i) of section 3.
Figure 1(a) shows the sub-Poissonian statistics for all the
values of $z$ and different values of $\eta=0.1,0.2$. As is
observed for the $z$ values smaller than $\simeq 9$ the
antibunching effect is stronger when $\eta$ is selected to be
0.2. This feature reverses when $z\geqslant9$. Figure 1(b) shows
squeezing in $x$-quadrature for both values of $\eta=0.1,0.2$. It
is seen that the squeezing is stronger for $\eta=0.2$ for all the
values calculated here. Interestingly, these states possess both
the antibunching and squeezing properties for all the range of
values of $z$. Figures 2(a) and (b) are related to the trapped ion
states in item (iii) of section 3 ($z = 0$). Figure 2(a) shows
squeezing in p-quadrature occurred for $\eta=0.1,0.2$ and all the
values of $\lambda$. It can be observed that there is no serious
sensitivity on the Lamb$-$Dicke parameter $\eta$. Figure 2(b) shows
the $Q$  parameter of the states in terms of $\lambda$. The
sub-Poissonian statistics will be evident about $\lambda
\leqslant 4$ . Figures 3(a) and (b) describe the properties of
the trapped ion intelligent states in item (iv) of section 3 (in
all of these plots we have set $z = 0.5$). Figure 3(a) shows the
super-Poissonian statistics for these states (or equivalently
bunching of the photons) for almost all the values of $\lambda$
when $\eta=0.2$, except in a small interval of $\lambda$ about
$\lambda \simeq 0.75$. Figure 3(b) shows the uncertainty in
p-quadrature as a function $\lambda$ parameter. It is seen that
squeezing occurred only when $\lambda$ is about 0.9 or greater
(with $\eta$ chosen as 0.2). As it may be expected, squeezing in
$p$ occurred for a wide range of values of $\lambda$ (with
$\eta=0.2$). Our calculations show that when $\lambda \geqslant
0.9$ squeezing is evident.\\

(2) \textit{Harmonious states}. Another system we will accomplish
in the present paper is the harmonious states characterized by the
nonlinearity function
 \begin{equation}\label{DSS1o1}
 f_{HS}(n)=\frac{1}{\sqrt{n}}.
 \end{equation}
 Keeping in mind relations (6) and (28), it can be observed that the obtained lowering operator
is equivalently the nonunitary Susskind$-$Glogower operator
$\exp(i \hat{\Phi}) = a(a^\dagger a)^{-1/2}$. It has been shown that
the probability operator measures generated by Susskind$-$Glogower
operator yield the maximum likelihood quantum phase estimation
for an arbitrary input state [27].

The nonlinear coherent states associated with the harmonious
states are restricted to a unit disc due to the requirement $0 < N
< \infty$. Clearly, this latter condition, i.e. the normalization
of the states, is also necessary for the intelligent states to
belong legally to the Hilbert space.

The following results are extracted from our numerical
calculations for the intelligent states of harmonious states. In
figures 4(a) and (b), we have plotted the quantum statistics of
the intelligent states of harmonious states in the form of item
(i) of section 3. Figure 4(a) demonstrates the super-Poissonian
exhibition for all the values of $z$. Figure 4(b) shows squeezing
in p-quadrature for all $z < 1$. It is seen that while $z$
increases, the squeezing will be stronger. Figures 5(a) and (b)
are related to the intelligent states of harmonious states if the
type of item (iii) of section 3 is considered ($z$ = 0). Figure
5(a) shows squeezing in p-quadrature versus $\lambda$, from which
the squeezing is readily seen for all the values of $\lambda$.
From the plot of p we observe that it has a maximum at $\simeq1$
and when $\lambda$ becomes greater or smaller than 1 the squeezing
will be stronger. When $\lambda$ becomes greater the squeezing
will be deeper. In figure 5(b), we have plotted the $Q$ parameter
as a function of $\lambda$, the super-Poissonian statistics
exhibition is clear from it. Figures 6(a)–(c) concerned with the
harmonious state's intelligent states in (iv) of section 3 ($z$ =
0.5). Distinctly, figure 6(a) shows the bunching of photons for
the corresponding intelligent states for a wide range of the
values of $\lambda$, except in a small interval of $\lambda
\simeq 0.75$. In figure 6(b), we have plotted the uncertainty in
p-quadrature in terms of $\lambda$ parameter. It is seen that
squeezing occurred only when $\lambda \geqslant 0.75$. Figure
6(c) shows the fluctuations of position against $\lambda$. As it
may be seen, squeezing in $x$ occurred when $0.2 \leqslant
\lambda \leqslant 0.75$.\\

(3) \textit{Hydrogen-like spectrum}. In the previous two
examples, we have considered that the systems with their
nonlinear coherent states were already known. To this end, we
want to apply our proposed formalism onto hydrogen-like spectrum,
a quantum system with known discrete spectrum:
\begin{equation}\label{ftyg}
   e_n=1-\frac{1}{(n+1)^2}.
   \end{equation}
The nonlinearity function in this case may be expressed as [24, 25]

\begin{equation}\label{temop1}
   f_H(n)=\frac{\sqrt{n+2}}{n+1}.
   \end{equation}

 The nonlinear coherent states associated with the hydrogen-like spectrum introduced
in (30) may be defined only on a unit disc of the complex plane.
In what follows, the nonclassicality nature of the three classes
of hydrogen-like's intelligent states has been discussed
numerically in detail. Figures 7(a) and (b) expressed the detail
statistics of hydrogen-like spectrum's intelligent states of item
(i) type in section 3. Figure 7(a) shows  the super-Poissonian
statistics of the states for all $z$, although very near $z \simeq
0$ the Poissonian statistics is observed. Figure 7(b) shows that
the p-quadrature is squeezed for all the values of $z$. As is
clear for the large $z$ the squeezing becomes stronger. Our
calculations show that squeezing can never be observed in
position operator, as the Heisenberg uncertainty predicts.
Figures 8(a) and (b) are related to hydrogen-like spectrum's
intelligent states in item (iii) of section 3 ($z = 0.5$). In
figure 8(a) squeezing is shown in p-quadrature for all the values
of $\lambda$. It is seen that $q_2$ has a peak about
$\lambda\simeq1$, which means that the squeezing has a minimum at
this point. Figure 8(b) shows the Mandel parameter of the hydrogen
intelligent states in terms of $\lambda$. The super-Poissonian
nature of these states and a minimum of about zero at
$\lambda\simeq1$ is visible (nearly Poissonian). In figures
9(a)–(d), we have plotted the squeezing in $x$,$p$ and $Q$
parameters of hydrogen-like intelligent states of (iv) type of
section 3 ($z = 0.5$). Figure 9(a) shows the uncertainty in
$x$-quadrature versus $\lambda$. It is seen that there is a small
range of $\lambda$ for which the squeezing can be occurred. In
figure 9(b), it is shown that the squeezing in p-quadrature may be
expected when $\lambda\geqslant0.8$. Figure 9(c) represents the
super-Poissonian statistics for these states (or equivalently
bunching of the photons) for almost all the values of $z$, except
in a small interval of $z$. Figure 9(d) again shows $Q$ as a
function of $\lambda$ but with $z$ chosen as 1.5 instead of 0.5 in
figure 9(c). The plot in figure 9(b) is quantitatively different
from that in figure 9(c), particularly the range of
nonclassicality increases essentially.

After all it is an obvious point that the presented formalism may
straightforwardly also be used to reproduce the intelligent
states of $SU(1, 1)$ group [6], the P\"{o}shl$-$Teller and infinite
well potential [9].


\section{Summary and conclusion}\label{examples}

  To summarize, after introducing the general algebraic formalism based on the nonlinear
coherent states method for the construction of the generalized
intelligent states, three new classes of the intelligent states
corresponding to harmonious states, center-of-mass motion of
trapped ion and hydrogen-like spectrum have been introduced
explicitly. Then, we investigated their nonclassicality features
through discussing the quantum statistical properties of the
obtained states. In particular, the role of the three various
representations of the intelligent states to reduce the quantum
fluctuations of the quadratures of the electromagnetic radiation
field and the sub-Poissonian statistics has been established in
detail. The application of the technique for obtaining the
intelligent states of Penson$-$Solomon type with nonlinearity
function $f_{PS} = q^{1-n}$ with $0 \leqslant q \leqslant 1$,
Barut$-$Girardello and Gilmore$-$Perelomov type of $SU(1, 1)$ group and
all sets of KPS (Klauder$-$Penson$-$Sixdeniers) coherent states (the
cases where the nonlinearities have been discussed in [24]) may
be constructed and discussed in future works. The presented
formalism in this paper can be considered as complementary of the
earlier works of El Kinani and Daoud [9], but it may be
recognized that the main aim of our presentation is to discuss
the nonclassicality nature of some new classes of the intelligent
states obtained through a rather different algebraic approach
namely the nonlinear coherent states method.\\

\begin{flushleft}
 {\bf Acknowledgments}\\
\end{flushleft}
\begin{flushleft}
One of the authors (MKT) would like to thank Professor R Roknizadeh for
his valuable comments and essentially for leading him to this
field of research. Also thanks to the referees for their valuable
recommendations which improve the quality of the paper.\\
\end{flushleft}


 \end{document}